\documentclass[english]{article}
\usepackage{amsmath}
\usepackage{amssymb}
\usepackage{cite}

\makeatletter

\usepackage{babel}

\begin{document}
\begin{titlepage}
\thispagestyle{empty}

\bigskip

\begin{center}
\noindent{\Large \textbf
{Localization of a Model With $U(1)$ Kinetic Gauge Mixing}}\\

\vspace{0,5cm}

\noindent{R.I. de Oliveira Junior${}^{a}$, M.O. Tahim${}^{b}$, G. Alencar${}^{a}$ and R.R. Landim${}^{a}$\footnote{e-mail: ivan@fisica.ufc.br}}

\vspace{0,5cm}
 
{\it ${}^a$Departamento de F\'{\i}sica, Universidade Federal do Cear\'{a}-
Caixa Postal 6030, Campus do Pici, 60455-760, Fortaleza, Cear\'{a}, Brazil. 
 
\vspace{0.2cm}
 }
 {\it ${}^b$Universidade Estadual do Cear\'a, Faculdade de Educa\c c\~ao, Ci\^encias e Letras do Sert\~ao Central- 
R. Epitácio Pessoa, 2554, 63.900-000  Quixad\'{a}, Cear\'{a},  Brazil.
 }

\end{center}

\vspace{0.3cm}

\begin{abstract}
In this paper we study the localization of a model with kinetic gauge  mixing on a thin membrane. The model we discuss is a theory for millicharged particles, proposed by Holdom in 1985, and that now is object of study in the LHC. We propose a geometrical coupling between the gauge fields, the Ricci scalar and the Ricci tensor. We show that it is possible to localize  such a model by regarding specific values for the coupling constants. We find the solutions for the two gauge fields and discuss the localization for scalar fields that appears naturally in the process. We show that not necessarilly the gauge and scalar fields are localized at the same time.
\end{abstract}
\end{titlepage}

\section{Introduction}

Theories with kinetic mixing are beyond the standard model(SM) and they have been studied in the last years. From the viewpoint of four dimensions it can be justified by the presence of a hidden sector \cite{Berezhiani:2008gi}. A sector of particles that would be a mirrow to the standard model. It is also important to dark matter models \cite{Chun:2010ve}, charge quantization \cite{CasteloFerreira:2005qh,CasteloFerreira:2005ej} and the existence of millicharges \cite{Berezhiani:2008gi,Gabrielli:2015hua,Ball:2016zrp,Batell:2005wa,Davidson:2000hf,Holdom:1985ag}. From the perspectives of extra dimensions there are several studies too. For example, in brane-antibrane models kinetic mixing of gauge fields is very natural in nonsuper-symmetric string set-ups \cite{Abel:2003ue}. Kinetic mixing in string theory have been computed in several models \cite{Rizzo:1998ut,Bullimore:2010aj,Dienes:1996zr,Babu:1997st,Abel:2008ai,Blumenhagen:2005ga,Benakli:2009mk,Arvanitaki:2009hb, Goodsell:2010ie,Goodsell:2009pi,Goodsell:2009xc }. The question we want to treat in this work is about the existence and localization of these models in membranes in the context of extra dimensions. We know that, in four dimensions, there are a bunch of models with two gauge factors. These models give explanations for some problems in Physics and even suggest new Physics. However, as we know, the fields in four dimensions must be localized, and none of theses models mention this fact. 

In the scenario of extra dimensions, the Randall Sundrum model(RS) \cite{Randall:1999ee,Randall:1999vf} achieved great success and, despite of it, contained a crucial problem: the localization of the  gauge fields. In that model they proposed that just gravity would be able to travel in a five dimensional bulk, bounded by four dimensional universes called branes, one with positive and other with negative tension, while the fields of the standard model would be trapped in theses 4-dimensional universes. However, they did not mention a mechanism of localization of fields other than gravity. 

Many works suggesting mechanisms to localize the gauge field have been proposed \cite{Kehagias:2000au,Zhao:2014iqa,Bajc:1999mh,Ghoroku:2001zu,Alencar:2014moa}. The essence of all of these mechanisms is to  modify the action of the electromagnetic field. In one of them it is made the suggestion that to localize the gauge field it is necessary to add a scalar field in the bulk \cite{Kehagias:2000au}. Another mechanism was proposed in \cite{Zhao:2014iqa}. In this last one, the authors assume that the 5-D gauge field has a dynamical mass term that is proportional to the Ricci scalar. These methods have problems that are fixed in \cite{Alencar:2014moa}. Where the authors use a non minimal coupling  to localize the gauge field. In this work we use the geometrical coupling  to localize a model with kinetic mixing of gauge fields. In order to do this, we couple the gauge field with the Ricci scalar \cite{Alencar:2014moa} in such a way that we can work without too many parameteres to fix. We also show that the coulpling with the Ricci tensor is an efficient method to localize the gauge fields. 

The organization of the paper follows.  In the second section, we propose a generalization of Holdom`s action \cite{Holdom:1985ag} to five dimensions and evaluate basic localization mechanisms. In the third section, we propose in fact a localization mechanism to Holdom`s model. In section four, we analysis the localization of our model using the Ricci tensor. In the fifth section, we treat the localization of the scalar components of the fields.  Finally, we present results, discussions and perspectives of future work.

\section{Two Gauge Fields in Five Dimensions With Minimal Coupling}

   In this section we  generalize to five dimensions the model presented in \cite{Holdom:1985ag} and try to localize it into a thin membrane. The generalized action reads:

\begin{eqnarray}
	S= -\int d^{5}x \sqrt{-g} \left[ \frac{1}{4} g^{MP} g^{NQ} F_{MN} ^{(1)} F_ {PQ}^{(1)} \right. \nonumber \\ \left.
	+ \frac{1}{4} g^{MP} g^{NQ} F_{MN} ^{(2)} F_ {PQ} ^ {(2)} + \frac{\lambda}{2} g^{MP} g^{NQ} F_{MN} ^{(1)} F_ {PQ}^{(2)} \right] \label{(11)} .
\end{eqnarray}	
Differently from \cite{Zhao:2014iqa}, we consider in this work the conformal (RS) metric;

\begin{equation}
	ds^{2} = e^{2A(y)} ( \eta_{\mu \nu} dx^{\mu} dx^{\nu} + dy^{2}) \label{(12)}. 
\end{equation}
Here we are regarding two gauge fields and, consequently, we have two field strengths. The uppercase latin indexes reffer to the 5-dimensional space-time. Varying the action (\ref{(11)}) with respect to $A_{M} ^{(1)}$ and to $ A_ {M} ^{(2)} $ , we obtain the following equations of motion:
\begin{equation}
	\partial_ {N} \left( \sqrt{-g} g^{MP}g^{NQ} F_ {PQ} ^{(1)} \right)  + \lambda \partial _ {N} \left( \sqrt{-g} g^{MP} g^{NQ} F_ {PQ} ^{(2)} \right) =0 \label{(13)}
\end{equation}

\begin{equation}
	\partial_ {N} \left( \sqrt{-g} g^{MP}g^{NQ} F_ {PQ} ^{(2)} \right)  + \lambda \partial _ {N} \left( \sqrt{-g} g^{MP} g^{NQ} F_ {PQ} ^{(1)} \right) =0. \label{(14)}
\end{equation}
Then, there is a discrete symmetry in this model that is given by the exchange of the two gauge fields $ A_ {M} ^{(1)} \leftrightharpoons	A_ {M} ^{(2)} $.

Using the equations (\ref{(13)}) and (\ref{(14)}), we obtain
\begin{equation}
	(1-\lambda^{2})  \partial_{N} \left( \sqrt{-g} g^{MP} g^{NQ} F_ {PQ} ^{(i)} \right) = 0, \label{(15)}
\end{equation}
where $i=1,2$.
Fixing $M=5$ in the equation (\ref{(15)}) we get

\begin{equation}
	\partial_ {\mu} \left( e^{A} F ^{5 \mu} _ {i} \right)=0 .\label{(16)}
\end{equation}
Now by fixing $M=\nu$ we find

\begin{equation}
	\partial_ {5} ( e^{A} F^{\nu 5} _ {(i)}) +  e ^{A} \partial_ {\mu} ( F^{\nu \mu} _ {(i)}) =0 . \label{(17)}
\end{equation}
Because we are working in the massless case, we can parametrize the 5-D gauge boson in the form $ A^{M} = (A^{\mu}, A^{5} = 0)$ , and use the Lorentz gauge $ \partial_{\mu} A^{\mu}= 0$. Using these facts in (\ref{(17)}) we get the equation:

\begin{equation}
	\partial_{5}( e^{A} \partial ^{5} A^{\nu}_ {(i)} ) + e^{A} \Box A^{\nu}_ {(i)} =0. \label{(18)}
\end{equation}

We propose a solution for the gauge field in the form: 

\begin{equation}
	A^{\nu} _ {(i)}(x,y) = \tilde {A^{\nu}_ {i}}(x)e ^{- \frac{A}{2}} \psi_ {i}(y). \label{(19)} 
\end{equation}
Putting (\ref{(19)}) in (\ref{(18)}) we obtain the following equations:
\begin{equation}
	\Box \tilde {A^{\nu}_ {i}}(x) = m^{2} \tilde{A}{^{\nu}_ {i} (x) } \label{(20)}
\end{equation}

and

\begin{equation}
	\psi''_ {i} - \left( \frac{A''}{2} + \frac{(A')^{2}}{4}\right) \psi_ {i}= - m^{2} \psi_ {i} . \label{(21)}
\end{equation}
For the zero mode we get the solution 
\begin{equation}
	\psi_{i} = e^{\frac{A}{2}}.\label{(22)}
\end{equation}
With this solution the zero mode of $ A^{\nu}_ {(i)}$  is not localized. And it is easy to be realized. If we put the value found for $\psi_{i}$ in (\ref{(19)}), we get

\begin{equation}
	A^{\nu}_{(i)}=\tilde{A}^{\nu}_ {(i)}.\label{(23)}
\end{equation}
With this, the effective theory becomes
\begin{eqnarray}
S=- \int_{-\infty}^{\infty}  e^{A} dy \int d^{4}x \left[ \frac{1}{4}\tilde{F}^{(1)}_{\mu\nu}\tilde{F}^{\mu\nu}_{(1)} +  \frac{1}{4}\tilde{F}^{(2)}_{\mu\nu}\tilde{F}^{\mu\nu}_{(2)} \right. \nonumber \\ \left.  +  \frac{\lambda}{2}\tilde{F}^{(1)}_{\mu\nu}\tilde{F}^{\mu\nu}_{(2)}  \right] .\label{(24)}
\end{eqnarray}

Once that we are working based in the Randall Sundrum model, we use its warp factor $ A= -\ln(k\left|y\right| + 1)$. Performing the integral that appears in front of the equation (\ref{(24)}), we get a not convergent result. In \cite{Batell:2005wa}, they get a confined theory, without adding any potential term, because they work in the Randall Sundrum type-I, where the extra dimension is compactified. As we are working with the RS-II, we need to add some potential terms to the action (\ref{(11)}), in order to localize the effective theory.

\section{Two Gauge Fields With Non-minimal Coupling  }
 We saw that the action presented in the last section is not a good candidate for an action for kinetic mixing of gauge fields in five dimensions, once that it does not localize the gauge fields as is proposed in \cite{Kehagias:2000au,Zhao:2014iqa,Bajc:1999mh,Ghoroku:2001zu,Alencar:2014moa}. We now modify this action taking advantage of the model with geometrical coupling:
 
 \begin{align}
 	 & S= -\int d^{5}x \sqrt{-g} \left[ \frac{1}{4} g^{MP} g^{NQ} F_{MN} ^{(1)} F_ {PQ}^{(1)} \right. & \nonumber \\ 
 	 & \left. + \frac{1}{4} g^{MP} g^{NQ} F_{MN} ^{(2)} F_ {PQ} ^ {(2)} +  \frac{\alpha}{2} g^{MP} g^{NQ} F_{MN} ^{(1)} F_ {PQ}^{(2)}  + \right. &  \\      
 	 & \left. \frac{\lambda_{1}}{2}R g^{MN} A_{M}^{(1)} A_{N} ^{(1)} +  \frac{\lambda_{2}}{2}R g^{MN} A_{M}^{(2)} A_{N} ^{(2)} + \lambda_{3}R g^{MN} A^{(1)}_ {M}A^{(2)}_ {N} \right] &  \nonumber \label{(25)}.
 \end{align}
 We have added the coupling between the Ricci scalar and the gauge fields through the last three terms above.
 Again we vary with respect to $ A_{M} ^{(1)}$ and to $ A_ {M} ^{(2)} $, and we get:
 \begin{eqnarray}
 	\partial_ {M} \left( \sqrt{-g} F^{MN} _{(1)} \right) + \alpha  \partial_ {M} \left( \sqrt{-g} F^{MN} _{(2)}\right) - \nonumber \\ \lambda_ {1} R \sqrt{-g} A^{N} _ {(1)} - \lambda_ {3} R \sqrt{-g} A^{N} _ {(2)} = 0\label{(26)}
 \end{eqnarray}
 and

 \begin{eqnarray}
  \partial_ {M} \left( \sqrt{-g} F^{MN} _{(2)} \right) + \alpha  \partial_ {M} \left( \sqrt{-g} F^{MN} _{(1)}\right) - \nonumber \\ \lambda_ {2} R \sqrt{-g} A^{N} _ {(2)} - \lambda_ {3} R \sqrt{-g} A^{N} _ {(1)} = 0. \label{(27)}
 \end{eqnarray}
 Note that there is a discrete symmetry in this model when we change $ \lambda_{1}$ for $ \lambda_{2}$ and $ A^{(1)}_ {M} $ for $  A^{(2)}_ {M} $.
 
 
 
 
 We need to do some manipulations in  the equations (\ref{(26)}) and (\ref{(27)}) using the antisymmetry of $ F_{MN}$, split the vectorial and scalar parts of the fields applying the identities between the field Strength and the gauge field found in \cite{Alencar:2014moa} and taking the gauge field as follow

 
 
 

 
 \begin{equation}
 	A^{M}= (A^{\mu}, A^{5}). \label{(36)}
 \end{equation}
 The four dimensional field is given by:
 \begin{equation}
 	A^{\mu} = A^{\mu} _ {T} + A^{\mu} _ {L} , \label{(37)}
 \end{equation}
 where $A^{\mu} _ {T} $ and $ A^{\mu} _ {L}$ are the transverse and the longitudinal parts of the field. Their explicit form is:
 
 \begin{equation}
 	A^{\mu} _ {T} = ( \delta ^{\mu}
 	_ {\nu} -  \frac{\partial^{\mu} \partial_ {\nu}}{\Box}) A^{\nu}   ~~~~~~~ A^{\mu} _ {L} = \frac{\partial^{\mu} \partial_ {\nu}}{\Box} A^{\nu} . \label{(38)}
 \end{equation}
 	After all this work, we finally get the following matricial form for the equations (\ref{(26)}) and (\ref{(27)})
 

 
 
 \begin{eqnarray}
 	\begin{pmatrix} 
 		\Box + e^{-A} \partial_{5}(e^{A}\partial^{5}) & 0   \\
 		0 &  \Box + e^{-A} \partial_{5}(e^{A}\partial^{5})  \\ 
 	\end{pmatrix} 
 	\begin{pmatrix} 
 		A^{\nu}_{T,(1)}   \\
 		A^{\nu}_{T,(2)} \\ 
 	\end{pmatrix} \nonumber \\
 	=-\frac{Re^{2A}}{(1-\alpha^{2})} 
 	\begin{pmatrix} 
 		(\alpha\lambda_{3} - \lambda_{1}) &  (\alpha\lambda_{2} -\lambda_{3})  \\
 		( \alpha\lambda_{1} - \lambda_{3})  & (\alpha\lambda_{3} - \lambda_{2})   \\ 
 	\end{pmatrix} 
 	\begin{pmatrix} 
 		A^{\nu}_{T,(1)}   \\
 		A^{\nu}_{T,(2)} \\ 
 	\end{pmatrix}. \label{(41)}
 \end{eqnarray}
 
 We have to diagonalize the matrix of the parameters in the equation above
in order to get separated equations. The matrix of the eigenvectors is given by:

 

  
 \begin{equation}
 P=
 \begin{pmatrix}
 \frac{\alpha \lambda_{2} -\lambda_ {3}}{\frac{\lambda_{1}-\lambda_{2}}{2} + \frac{1}{2}B} &  \frac{\alpha \lambda_{2} -\lambda_ {3}}{\frac{\lambda_{1}-\lambda_{2}}{2} - \frac{1}{2}B} \\
 1 & 1 \\
 \end{pmatrix}, \label {(44)}
 \end{equation}
 where $B=\sqrt{(\lambda_{1}-\lambda_{2})^{2} + 4\lambda_{3}[\lambda_{3} -\alpha(\lambda_{1} + \lambda_{2})] + 4\alpha^{2} \lambda_{1} \lambda_{2}}$.
After redefining the fields as follow 
  \begin{eqnarray}
 	\begin{pmatrix}
 		\bar{A}^{\nu}_ {T,(1)}\\
 		\bar{A}^{\nu}_ {T,(2)} \\       
 	\end{pmatrix} \nonumber
 	= P^{-1}
 	\begin{pmatrix}
 		{A}^{\nu}_ {T,(1)}\\
 		{A}^{\nu}_ {T,(2)} \\
 	\end{pmatrix}, \label{(45)}
 \end{eqnarray}
and making some mathematical steps we get the desired separated equations:
 
 
 \begin{eqnarray}
 	\Box \bar{A}^{\nu}_ {(i),T} + e^{-A} \partial_{5}[ e^{A} \partial^{5} \bar{A}^{\nu}_ {(i),T}] &=& \nonumber \\ \frac{-R}{(1 - \alpha^{2})} e^{2A}  \bar{A}^{\nu} _ 
 	{(i),T}\left(\alpha \lambda_{3} - \frac{(A\mp B)}{2}\right).  \label{(48)}
 \end{eqnarray}
 
  Doing the Kaluza-Klein decomposition $ \bar{A} ^{\nu} _ {(i),T} = e^{-\frac{A}{2}} \tilde{A} ^{\nu} _ {(i),T} \psi_{i}(y) $ , where $i=1,2$ in equations (\ref{(48)}) , we obtain 
  
  \begin{equation}
   \psi''_{(i)} -\left\{ A''\left[ \frac{1}{2} + \frac{8}{(1-\alpha^{2})}\left(\alpha \lambda_{3} - \frac{(A\mp B)}{2}\right)\right] + A'^{2}  \left[ \frac{1}{4} + \frac{12}{(1-\alpha^{2})} \left( \alpha \lambda_{3} - \frac{(A\mp B)}{2} \right)\right] \right\}\psi_{(i)}. \label{(50)}
  \end{equation}
From these equations we get two relations, for $i=1,2$, that localize $ A^{\nu}_{T,(1)}$ and $A^{\nu}_{T,(2)} $, respectively. They are

\begin{equation}
 \left(\alpha \lambda_{3} - \frac{(A-B)}{2}\right)= \frac{(1-\alpha^{2})}{16}~~~~  ,~~~~  \left(\alpha \lambda_{3} - \frac{(A+B)}{2}\right)= \frac{(1-\alpha^{2})}{16}. \label{(51)} 
\end{equation}
Adding the two relations in (\ref{(51)}), we obtain that $B=0$. If $B=0$ the matrix of the eigenvectors (\ref{(44)}) does not have inverse, and we can not diagonalize the matrix of the coupling constants in (\ref{(41)}).That way we are not allowed to work with any value for the coupling constants. We must find a relation for them that obeys the localization condition $B=0$, and allow the matrix (\ref{(44)}) to have an inverse. Let us see wich relation is that. 

Writing (\ref{(44)}) as

\begin{equation}
 P= 
 \begin{pmatrix}
  K_{1} & K_{2} \\
  1 & 1 \\
 \end{pmatrix}. \label{(52)}
\end{equation}
Where $K_{1}$ and $K_{2}$ are given by
\begin{equation}
 K_{1}= \frac {2(\alpha \lambda_{2} - \lambda_{3})}{\lambda_{1} -\lambda_{2} + B} , ~~~~  K_{2}= \frac {2(\alpha \lambda_{2} - \lambda_{3})}{\lambda_{1} -\lambda_{2} - B}. \label{(53)} 
\end{equation}
Dividing $K_{1}$ by $K_{2}$, and making $ C= \lambda_{1} - \lambda_{2}$, we get $(K_{1} - K_{2}) C =0$. As $K_{1}$  must be different of $K_{2}$, in order to have $detP \neq 0$, we end up with $C=0$. This leads to $\lambda_{1}=\lambda_{2}=\lambda$. Substituing $\lambda_{1}=\lambda_{2}$ in the expression for $B$ and equalizing to zero, we find that $\lambda_{3}= \lambda \alpha$. With these values for the parameters, the matrix for these constants in (\ref{(41)}) gets a diagonal form. Let us now work with the separated equations that comes from (\ref{(41)}).

By putting the allowed values for the coupling constants we get

\begin{eqnarray}
 	\Box A^{\nu}_ {(i),T} + e^{-A} \partial_{5}[ e^{A} \partial^{5} A^{\nu}_ {(i),T}]= \lambda Re^{2A} A^{\nu}_{T,(i)} . \label{54}
 \end{eqnarray}
Using the ansatz for the solutions as before, $ {A} ^{\nu} _ {(i),T} = e^{-\frac{A}{2}} \tilde{A} ^{\nu} _ {(i),T} \psi_{i}(y) $,  and the Ricci scalar computed from the conformal Randall-Sundrum metric $ R= -e^{-2A}( 8A'' +12 {A'}^{2}) $ we obtain from the equation (\ref{54})

 \begin{eqnarray}
 \psi^{''}_ {i} - \left[A'^{2} \left(\frac{1}{4} -12 \lambda \right) + A''\left( \frac{1}{2} - 8\lambda\
  \right) \right] \psi_i = 0. \label{57}
 \end{eqnarray}
 For the zero mode, which means $m=0$, we can propose a solution of the form $ \psi_{i} = e^{cA}$. With this we get the relations: 
 
 \begin{equation}
 \lambda = 0    ,  ~~~~~  \lambda= -\frac{1}{16}.  \label{58}
 \end{equation} 
 For the first one we get one solution, $\psi_i = e^{\frac{A}{2}}$, that does not localize the fields, while for the second one we have $\psi_i = e^{A}$, that localize the fields. Using the solution that localizes the fields we get the final form for the two gauge vectors

 \begin{eqnarray}
 A^{\nu}_{(i),T}= e^{\frac{A}{2}} \tilde{A} ^{\nu}_{T,i}. \label{(64)}
 \end{eqnarray}
 
 We then have the two gauge fields. One can be interpreted as the usual gauge field and other the gauge field from the hidden sector \cite{Berezhiani:2008gi}. What we can not tell yet is which one is from the hidden sector . It is an interesting fact, that both fields have the same form and the same coupling constant with the Ricci scalar. If, for example, we say that one of the fields in (\ref{(64)}) is the usual $U(1)$ gauge field, interaction of charged particles from the hidden sector with this field can generate millicharged particles (particles with fraction of the electron charge). Yet, interaction of massive fermions with the other gauge field in (\ref{(64)}) might create dark matter \cite{Chun:2010ve}.   We see that with the geometrical coupling it is possible to localize a model with two gauge fields. We will see it clearly  when we show the effective action.

We are now going to show how is the effective action. In order to show it we have to decompose all terms in the starting action by using our separation for the gauge field $ A^{\mu} = A^{\mu}_{T} + A^{\mu}_{L}$, the condition $ \partial_{\mu}A^{\mu}_{T}=0$ and the fact that part of the component $ \frac{1}{2}g^{55}g^{\mu\nu}F_{\mu5}F_{\nu5}$ cancels out with the components that are coupled with the Ricci scalar. After doing all of that we see that we can separate the action  in one vectorial part and in a scalar part:

\begin{eqnarray}
S_{V} =-\int d^{5}x \sqrt{-g}\left[ \frac{1}{4}g^{\mu\nu}g^{\lambda\rho} F^{(1),T}_{\mu\lambda}F^{(1),T}_{\nu\rho} \right. \nonumber \\ \left. +\frac{1}{4}g^{\mu\nu}g^{\lambda\rho} F^{(2),T}_{\mu\lambda}F^{(2),T}_{\nu\rho}   + \frac{\alpha}{2}g^{\mu\nu}g^{\lambda\rho} F^{(1),T}_{\mu\lambda}F^{(2),T}_{\nu\rho}\right] \label{(67)}
\end{eqnarray}
\begin{eqnarray}
S_{S} = -\int d^{5}x \sqrt{-g}\left[ \frac{1}{2}g^{55}g^{\mu\nu}\partial_{\mu}A^{(i)}_{5}\partial_{\nu}A^{(i)}_{5} + \right.\nonumber \\ \left. \frac{1}{2}g^{55}g^{\mu\nu}A_{5}^{(i)} \partial_{5}\partial_{\mu}A^{(i)}_{\nu,L} + \frac{1}{2}g^{55}g^{\mu\nu}\partial_{5}(\partial_{\nu}A^{(i)}_{\mu,L})A^{(i)}_{5} +  \right.\nonumber \\ \left. \frac{1}{2}g^{55}g^{\mu\nu}\partial_{5}A^{(i)}_{\mu,L} \partial_{5}A^{(i)}_{\nu,L} + \alpha g^{55}g^{\mu\nu}\partial_{5}A^{(1)}_{\mu,L} \partial_{5}A^{(2)}_{\nu,L} \right.\nonumber \\ \left. \alpha g^{55}g^{\mu\nu} \partial_{5} \partial_{\nu}A^{(1)}_{\mu,L}A^{(2)}_{5} + \alpha g^{55}g^{\mu\nu}A^{(1)}_{5} \right.\nonumber \\ \left. \partial_{5}\partial_{\mu}A^{(2)}_{\nu,L} + \alpha g^{55}g^{\mu\nu} \partial_{\mu}A^{(1)}_{5}\partial_{\nu}A^{(2)}_{5} + \right.\nonumber \\ \left. + \frac{\lambda_{i}R}{2}g^{55}A^{(i)}_{5} A^{(i)}_{5}  + \lambda_{3} R g^{55}A^{(1)}_{5} A^{(2)}_{5}   \right]. \label{(68)}
\end{eqnarray}
Now replacing the gauge fields that we found (\ref{(64)}) in the vectorial action (\ref{(67)}), we obtain:

\begin{eqnarray}
S_{V} = -\int_{-\infty} ^{\infty} e^ {2A} dy\int d^{4} x \left[ \frac{1}{4} \tilde{F}^{(1)}_ {\mu\lambda}\tilde{F}^{\mu\lambda}_{(1)} + \frac{1}{4} \tilde{F}^{(2)}_ {\mu\lambda} \right. \nonumber \\ \left. \tilde{F}^{\mu\lambda}_{(2)} + \frac{\alpha}{2} \tilde{F}^{(1)}_ {\mu\lambda} \tilde{F}^{(2)}_ {\mu\lambda}  \right]. \label{69}
\end{eqnarray}
Solving the integral $ \int ^{\infty} _{-\infty} e^{2A}dy$ with $ A(y)= - \ln( \left| y \right|  +1 )$ we get as result $\frac{2}{k}$. This means that the whole action above is localized. This action is an action for millicharged particles. And we showed that it can be localized using the geometrical coupling.
 

\section{Coupling with the Ricci tensor}

We also constructed a model for localization of the gauge fields using the Ricci tensor. In this attempt we used the Ricci tensor computed from the conformal RS metric that can be found at \cite{Alencar:2018cbk}. We proposed an action that is very similar to the one with the Ricci scalar. However, the Ricci tensor takes the place of the Ricci scalar. The actions reads:

\begin{eqnarray}
S= -\int d^{5}x \sqrt{-g} \left[ \frac{1}{4} g^{MP}g^{NQ}F^{(1)}_ {MN}F^{(1)}_{PQ} + \frac{1}{4} g^{MP}g^{NQ} F^{(2)}_ {MN}F^{(2)}_{PQ} \right. \nonumber \\ \left.  + \frac{\alpha}{2}g^{MP}g^{NQ}F^{(1)}_ {MN}F^{(2)}_{PQ} + \frac{\lambda_{1}}{2}R_{MN}g^{MP}g^{NQ} A^{(1)}_{P}A^{(1)}_{Q} \right. \nonumber \\ \left. + \frac{\lambda_{2}}{2}R_{MN}g^{MP}g^{NQ} A^{(2)}_{P}A^{(2)}_{Q} + \lambda_{3} R_{MN}g^{MP}g^{NQ} A^{(1)}_{P}A^{(2)}_{Q} \right]. \label{(70)}
\end{eqnarray}

Doing the same processes above where we vary the action with respect to  the two fields, combine the equations of motions, separate the scalar and vectorial parts, using the identities introduced in \cite{Alencar:2014moa} and put the resulting equations in matricial form we find:

\begin{eqnarray}
\begin{pmatrix} 
\Box + e^{-A} \partial_{5}(e^{A}\partial^{5}) & 0   \\
0 &  \Box + e^{-A} \partial_{5}(e^{A}\partial^{5})  \\ 
\end{pmatrix} 
\begin{pmatrix} 
A^{\nu}_{T,(1)}   \\
A^{\nu}_{T,(2)} \\ 
\end{pmatrix} \nonumber \\
=-\frac{R^{\nu\mu}}{(1-\alpha^{2})} 
\begin{pmatrix} 
(\alpha\lambda_{3} - \lambda_{1}) &  (\alpha\lambda_{2} -\lambda_{3})  \\
( \alpha\lambda_{1} - \lambda_{3})  & (\alpha\lambda_{3} - \lambda_{2})   \\ 
\end{pmatrix} 
\begin{pmatrix} 
A_{\mu , T}^{(1)}   \\
A_{\mu, T}^{(2)} \\ 
\end{pmatrix}. \label{(71)}
\end{eqnarray}

As we can see from (\ref{(71)}), we get an equation with structure similar to (\ref{(41)}) , what was already expected, once that we just replaced the Ricci scalar by the Ricci tensor . According to the process done in the previous section we have to have specific values for the parameters in order to localize both fields. With these parameters placed in (\ref{(71)}) we have two separated equations.

\begin{eqnarray}
\Box A^{\nu} _ {T, (i)} + e^{-A}\partial_{5} (e^{A} \partial^{5}A^{\nu}_{T,(i)}) = \lambda R^{\mu\nu}A^{T}_{\mu,(i)}. \label{(72)}
\end{eqnarray}
Here we have used the Ricci tensor computed from \cite{Alencar:2018cbk} with $D=5$. After some mathematical procedures, we get  the two Schrodinger type equations, with $m=0$. This time, when we impose that both gauge fields have to be localized,  we find the following relations for the parameters



\begin{equation}
\lambda=0 ,~~~~~  \lambda=-2 . \label{(76)}
\end{equation}
For $\lambda=0$ we get one solution for the Schrodinger equation that does not localize the fields. While that for $\lambda=-2$ we have $\psi_{(i)}= e^{\frac{5A}{2}}$, wich indeed localizes the fields. For this solutions we get a convergent result for the integral $\int^{\infty}_{-\infty} \psi_{i}^{2}dy$. We see that the coupling with the Ricci tensor is also a good way to localize the gauge fields.

\section{Scalar Action }
We now focus our attention in the scalar componentes for the coupling of the gauge fields with the Ricci scalar. We could do this analysis from the action (\ref{(68)}). However, for simplicity , we will use the components $A_{5}$ of  the equations (\ref{(26)}) and (\ref{(27)}):  

\begin{eqnarray}
 \partial_ {\mu}F ^{5 \mu} _ {(1)} 
+ \frac{R e ^{2A}}{(1-\alpha^{2})} (\alpha \lambda_{2} \Phi_{2}  + \alpha \lambda_ {3} \Phi_ {1} -\lambda_{1} \Phi_ {1} - \lambda_{3} \Phi_ {2} ) =0 \label{78}
\end{eqnarray}

and 
\begin{eqnarray}
\partial_ {\mu}F ^{5 \mu} _ {(2)} 
+ \frac{R e ^{2A}}{(1-\alpha^{2})} (\alpha \lambda_{1} \Phi_{1}  + \alpha \lambda_ {3} \Phi_ {2} -\lambda_{2} \Phi_ {2} - \lambda_{3} \Phi_ {1} ) =0. \label{79}
\end{eqnarray}
In the equations above we have done $A^{5}_{i}=\Phi_{i}$ . Considering the relations between the parameters, we can simplify this pair of equations in the following way:

\begin{eqnarray}
\boxdot\Phi_{i} + \partial^{5}[R^{-1} e^{-3A} \partial_{5} ( Re^{3A}\Phi_{i})] - \lambda Re^{2A} \Phi_ {i}=0. \label{80}
\end{eqnarray}
Here we have used the definition of the  field strength, made $i=1,2$ and the relation:

\begin{eqnarray}
\partial_ {\mu} A^{\mu}_ {i} = - R ^{-1} e ^{-3A}\partial_{5}(Re^{3A}\Phi_{i}). \label{81}
\end{eqnarray}
Using the method of separation of variables, with $ \Phi_ {i}(x,z)=\psi_ {i} (z) \phi_{i}(x), $ we get:

\begin{eqnarray}
\partial^{5} [R^{-1} e^{-3A} \partial_{5}(Re^{3A}\psi_{i}(z))] -\lambda R e^{2A} \psi_{i}(z)=- m^{2} \psi_{i}(z). \label{82}
\end{eqnarray}

This equation is similar to that one found in \cite{Alencar:2014moa}. It is our intention to put this equation in Schrodinger form. First, if we do the transformations $ g= e^{3A}$ and $ h=- \lambda e^{-A} g$ , we can write the equation above as follow:

\begin{eqnarray}
 -\psi''_ {i} - (\ln fg^{2})'\psi_{i}' - [(fg')'+ h]\psi= m^{2} (fg)^{-1} \psi_ {i}. \label{83} 
\end{eqnarray}
Comparing this equation with an identity found in \cite{Alencar:2012en}, we have, $ P =\ln (fg^{2})$, $V=-[(fg')'+h ]$ and $Q=(fg^{-1}) $.The potential for the Schrodinger type equation, in its general form, is given by: 


\begin{equation}
U(z) = \frac{V(y)}{\Theta(y)^{2}} + \frac{(P'(y)\Omega'(y)-\Omega''(y))}{\Omega{y}\Theta(y)^{2}}. \nonumber \\
\end{equation}
The first and second derivatives of $\Omega$ are:
\begin{equation}
\Omega'= [\ln(f^{-\frac{1}{4}} g ^{ -\frac{3}{4}} )]'\Omega~~~~, ~~~~  \Omega= f^{\frac{1}{4}}g^{\frac{3}{4}}(f^{-\frac{1}{4}}g^{-\frac{3}{4}})''\Omega. \nonumber \\
\end{equation}
With these derivatives we get the following form for the potential:

\begin{eqnarray}
U(z) = -[ (f g')' + h ]fg + 
\left[(\ln f^{-1} g^{-2})'(\ln f^{-\frac{1}{4}g ^{-\frac{3}{4}}})'\right.\nonumber \\ \left. -f^{\frac{1}{4}}g^{\frac{3}{4}}(f^{-\frac{1}{4}}g^{-\frac{3}{4}})\right]fg.\label{84}
\end{eqnarray}
In our case, we have set $ f=g^{-1}$  then we get for the potential:
\begin{equation}
U(z)= \frac{3}{4}\frac{{g'}^{2}}{g^{2}} -\frac{1}{2} \frac{g''}{g} - \lambda e^{-A} g \label{85}
\end{equation}

or

\begin{eqnarray}
U(z) = \frac{1}{4} (3A'+ (\ln R)')^{2} - \frac{1}{2} (3A'' + (\ln R)'') -\lambda R e^{2A}. \label{86}
\end{eqnarray}

If we consider an asymptotic RS metric, where  $\lim_{z \rightarrow \infty}=-20\kappa^{2} $, we get an asymptotic potential 

\begin{equation}
U= {A'}^{2} \left( \frac{9}{4} + 12 \lambda\right) -A''\left(\frac{3}{2} - 8\lambda \right). \label{87}
\end{equation}
The solution that localizes the zero mode of the scalar field is $\psi_{i}= e^{3A}$, and the value of the parameter is $ \lambda=\frac{9}{16}$. We can easily see that like in \cite{Alencar:2014moa}, the values for the parameters that localize the vectorial and scalar sectors are different. This allows us to say that we can not localize both sectors at the same time. A discussion about the simultaneous localization  of both sectors is made in \cite{Freitas:2018iil}.

\section{Conclusions}

In this paper, we proposed a localization for a model with $U(1)$  kinetic gauge mixing. In the beginning, we tried just generalizing to five dimensions the action proposed in \cite{Holdom:1985ag}. The way we found to get rid of this problem was trying the coupling of the gauge $U(1)$ fields with the Ricci scalar. In fact, it was a good solution for the problem. We showed that just for the following relation between the parameters: $\lambda_{1} = \lambda_{2}=\lambda$ and $\lambda_{3}=\lambda \alpha$, it is possible to localize the two vector field at the same time.The parameter $\alpha$ is free and it is responsible to give the strength of the kinetic gauge mixing, and make appear a particle with charge smaller than the electric charge. We also found the explicit solutions for both gauge fields in the localization mechanism.

We also tried the coupling with the Ricci tensor. Fortunately, it also was a good way of localizing our model. Of course it has some few differences from the previuos model. The solution and the parameter that localize the fields in that case are, respectively $ \psi_{i}= e^{A}$ and $ \lambda=-\frac{1}{16}$, while for the model with the Ricci tensor are $\psi_ {i}= e^{\frac{5A}{2}}$and $ \lambda=-2$.

The discussion about the localization of the scalar components of the fields was also made. To do that we considered the model with the Ricci scalar. Again we found different values for the solution and the parameter that localize these scalar components, they are $ \psi_{i}= e^{3A}$ and $\lambda= \frac{9}{16}$. As the values that localize the vectorial and scalar sectors are different, they can not be localized at the same time. The localization at the same time, of both sectors, are discussed in \cite{Freitas:2018iil}. All of these results  were only possible due to the power of the geometrical coupling apllied in this work.

We hope these results can help in understanding more about millicharges from the viewpoint of extra dimensions. As perspective of future work we intend to calculate the massive modes of this model as well as resonance oscillations. Yet, we want to see the correction in the coulomb's law due to the mixing of the gauge fields.

\section*{Acknowledgments}

The authors would like to thanks Alexandra Elbakyan and sci-hub, for removing all barriers
in the way of science. We acknowledge the financial support provided by the Conselho
Nacional de Desenvolvimento Cient\'ifico e Tecnol\'ogico (CNPq) and Funda\c{c}ao Cearense de
Apoio ao Desenvolvimento Cient\'ifico e Tecnol\'ogico (FUNCAP) through PRONEM PNE0112- 00085.01.00/16

\end{document}